\documentclass[a4paper,fleqn,usenatbib]{mnras}
\usepackage{newtxtext,newtxmath}
\usepackage[T1]{fontenc}
\usepackage{ae,aecompl}
\usepackage{graphicx}	
\usepackage{amsmath}	
\usepackage{amssymb}	
\usepackage{subfigure}
\usepackage{fix-cm}
\usepackage{txfonts}
\usepackage[utf8]{inputenc}
\usepackage{tablefootnote}

\title[Variability in the massive O+B Binary HD\,165246]{Detection of intrinsic
  variability in the eclipsing massive main sequence O+B binary HD\,165246}

\author[Johnston et al.]{
C. Johnston$^{1}$\thanks{E-mail: colecampbell.johnston@kuleuven.be},
B. Buysschaert$^{1,2}$,
A. Tkachenko$^{1}$,
C. Aerts$^{1,3}$,
and C. Neiner$^{2}$
\\
$^{1}$Instituut voor Sterrenkunde, KU Leuven, Celestijnenlaan 200D, 3001 Leuven, Belgium\\
$^{2}$LESIA, Observatoire de Paris, PSL Research University, CNRS, Sorbonne
Universit\'es, UPMC Univ.\ Paris 6, 
Univ.\ Paris Diderot, \\Sorbonne Paris Cit\'e, 5 Place Jules Janssen, 92195, Meudon, France\\
$^{3}$Dept. of Astrophysics, IMAPP, Radboud University Nijmegen, 6500 GL, Nijmegen, The Netherlands\\
}

\date{Accepted XXX. Received YYY; in original form ZZZ}

\pubyear{2017}

\begin{document}
\label{firstpage}
\pagerange{\pageref{firstpage}--\pageref{lastpage}}
\maketitle

\begin{abstract}
  We present the analysis of 29.77 days of \textit{K2\/} space photometry of the well-detached
  massive 4.6\,d O+B binary HD\,165246 (V=7.8) obtained during Campaign 9b. This analysis
  reveals intrinsic variability in the residual lightcurve after subtraction of 
  the binary model, in the frequency range $[0,10]$\,d$^{-1}$. This makes
  HD\,165246 only the second  O+B eclipsing binary with asteroseismic potential. 
  While some of the frequencies are connected with the rotation of the primary,
   others are interpreted as due to oscillations with periodicities of order 
   days. The frequency resolution of the 
  current dataset does not allow to distinguish between frequencies due to standing
  coherent oscillation modes or travelling waves. Future time-resolved
  high-precision spectroscopy covering several binary orbits will reveal whether
  HD\,165246 is a Rosetta stone for synergistic binary and seismic modelling of
  an O-type star.
\end{abstract}



\section{Introduction}
The past decade has brought revolutionary advances in the fields of
asteroseismology and stellar evolution for intermediate- to low-mass
Main-Sequence (MS) and Red Giant (RG) stars given the high-precision, high
duty-cycle lightcurves assembled by the space-based CoRoT \citep{corot2009} and
nominal \textit{Kepler} missions \citep{kepler2010}. Some of the most significant advances
for intermediate- (A-F) to high-mass (O-B) MS stars include: i) determination of
surface-to-core rotation profiles
\citep{kurtz2014,saio2015,murphy2016,schmid2016}, ii) identification of regular
g-mode period series and their use to constrain rotation profiles
\citep{vanreeth2015,papics2017,triana2015}, and iii) calibration of chemical
mixing caused by rotation and overshooting
\citep{moravveji2016,vanreeth2016}. In addition, \textit{Kepler} delivered a few
thousand eclipsing binaries, which allow for largely model-independent estimates
of stellar parameters. The co-evol nature of binaries provide the only other calibration for internal physics, such as mixing and overshooting, by forcing
model evaluations at the same ages against spectroscopy, as was shown by
\cite{claret2016}. In the cases where pulsations are observed in eclipsing
binaries, asteroseismic and binary modelling produce complementary parameter
estimates, enabling synergistic modelling from two independent methods. These
systems are of high astrophysical importance due to their unique ability to
constrain internal physics to a high precision. In unique cases, binary systems 
with a high eccentricity have been shown to exhibit periastron brightening and tidally excited pulsations, e.g. \cite{welsh2011,hambleton2016}. \cite{pablo2017} recently reported the first case of an eccentric O-type binary exhibiting tidally induced pulsations with the BRITE-constellation satellites, although the system does not show eclipses.

Due to a scarcity of O-type stars in the CoRoT fields, and an altogether lack
thereof in the nominal \textit{Kepler\/} field, these stars have not been
analysed in such detail as their lower-mass counterparts. This has left models
of these stars' interiors uncalibrated by asteroseismology or high-precision
binary modelling, and hence has not allowed to evaluate physical mechanisms such
as rotational mixing and core-overshooting in massive stars. On the other hand,
Internal Gravity Waves (IGW) generated at convective/radiative boundaries inside
massive stars which transport chemicals and angular momentum
\citep[e.g.,][]{rogers2013,alvan2015} have recently been detected in
high-precision space data \citep{aerts2015,aerts2017}. However, while seismic
analysis of coherent oscillation modes compares theoretically computed 
frequencies for a given stellar model against observationally determined frequencies,
no direct method of comparison has yet been developed for analysis of IGWs. Instead, 
observations of IGWs are compared with computationally expensive 2D and 3D simulations
which require a fixed mass and age ranges as input.

After it's re-purposing, the \textit{K2} mission began observing multiple fields,
 providing the opportunity for high-quality space-based observations of massive
stars, such as HD\,165246. Originally
 characterized as an O8 V star by \cite{walborn1972}, HD\,165246 (EPIC
224466741) was later discovered to be an eclipsing binary by \cite{otero2007},
making this object a prime candidate for binary modelling to constrain interior
physics. \cite{mayer2013} assembled spectroscopy alongside 617 photometric ASAS
observations taken in the V-Band to model the binarity of the system with the
PHOEBE binary modelling code \citep{prsa2005}. Their analysis rendered a
complete orbital solution with a 4.6\,d period assuming no third light, revealing a slight
eccentricity, super-synchronous rotation in the hot primary, and an argument of
periastron near 90 degrees. However, the authors noted caution at the value of
the eccentricity and suggested that it should be investigated further with
higher quality data.

\cite{sana2014} utilized interferometric observations obtained with the Sparse
Aperture Masking (SAM) mode of the NACO instrument on the Very Large Telescope at ESO
to deduce that the 4.6\,d binary studied by \cite{mayer2013}, denoted as HD\,165246Aa+Ab, is a member of a sextuple system with the companions being ~30 mas (HD\,165246B), ~1.9 arcsec (HD\,165246C), ~6.6 arcsec (HD\,165246D), and
~7.9 arcsec (HD\,165246E) away. Furthermore, \cite{sana2014} reported magnitude contrasts in
the H band, revealing that the closest two members to the inner binary (B+C) could
provide anywhere from 7.2 to 43.35\%  (within $1-\sigma$) of the total light
observed from the system, with the large uncertainties arising from the tertiary
being located at the detector limit of the NACO instrument. 

In the current analysis, we report the detection of intrinsic variability in the
\textit{K2\/} lightcurve of HD\,165246, implying that this object may become a
Rosetta stone for future asteroseismic calibration of the class of massive MS
O-type stars. Furthermore, the detection of this eclipsing system provides different
 constraints than the $\iota$ Ori system \citep{pablo2017} which uses
  the periastron brightening event to determine the system inclination.
\section{\textit{K2\/} Photometry}
\begin{figure}
	\centering
	\includegraphics[width=\columnwidth,keepaspectratio]{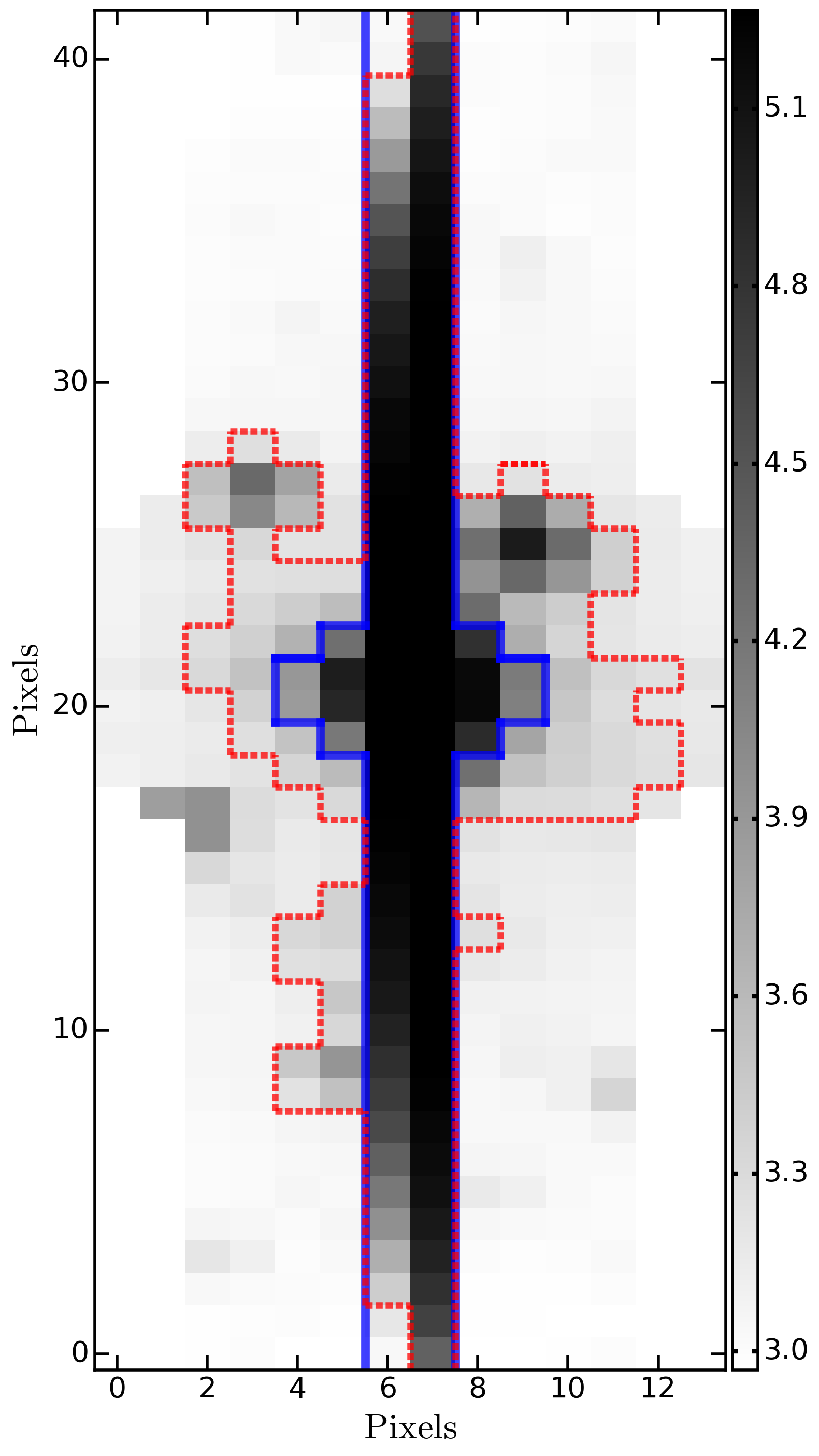}
        \caption{Comparison of the original mask (red) and our custom mask
          (blue) used to extract the \textit{K2} lightcurve. The grey scale indicates the
          logarithm of the co-added flux of all images (in e/s). Pixels are $3.96\times3.96$ 			      arcsec each.}
    \label{fig:reductions}
\end{figure}
HD\,165246 was observed in both the long-cadence (LC, 29.42\,min integration
time) and short-cadence (SC, 58.85\,s integration time) modes from 21 April to 1
July 2016 by the re-purposed \textit{K2\/} space satellite as a part of proposal
GO9913. \textit{K2\/} \citep{howell2014} is the continuation of the nominal
\textit{Kepler\/} mission which observed a single field between Cygnus and Lyrae
for nearly four years with $\mu$mag precision before the loss of two reaction
wheels. Although \textit{K2\/} suffers from a loss in precision and
observational time-base, it observes more fields, including those containing
massive stars. 

Community efforts to reduce \textit{K2\/} data have been made, resulting in a
host page at the Mikulski Archive for Space Telescope\footnote{MAST; {\tt
    https://archive.stsci.edu/k2/}} for the data extracted according to
\cite{vanderburg2014}. Inspection of the target pixel masks reduced according to
\cite{vanderburg2014} reveals that multiple contaminating sources were included
in the original mask of HD165246 (Fig.\,\ref{fig:reductions}).  To minimize contamination,
we downloaded the original target pixel data and performed a custom reduction
according to \cite{buysschaert2015}, avoiding the contaminating sources included
in the original mask. A comparison of the two masks can be seen in
Fig.\,\ref{fig:reductions} and illustrates why our reduction is more appropriate
to study the binary itself. It is important to note that all six members of the 
sextuple system are covered by just two of the large 3.96x3.96 arcsec \textit{Kepler} pixels.
 Our reduction resulted in nearly 30 days of
high-quality science data for which instrumental effects are negligible (some
1400 data points). As this is a bright target \citep[V=7.8,][]{ducati2002}, it
saturates the \textit{Kepler\/} pixels, smearing the signal across several of
them. This allows us to check that the binary signal and other variability
originate from the same source by varying the pixel mask and comparing the
resulting lightcurves. As such, we find that the variability remains in the
lightcurve despite quite drastic variations in mask selection for data
extraction, always avoiding the contaminating sources.

\section{Lightcurve Analysis}
Initial inspection of the lightcurve reveals a moderate reflection effect and non-sphericity of the primary given the increasing flux towards secondary eclipse, flat
bottom eclipses, and apparent variable signal present in both primary and
secondary eclipses, suggesting an intrinsically variable primary. Meaningful
investigation of the variability requires a two-step light-curve modelling
approach by which we first fit the binary orbit, remove it, and then examine the
residuals via time-series analysis.

\subsection{Binary Modelling}
\begin{table}
	\centering
	\caption{Varied and derived parameters for PHOEBE  modelling, adopting
          the parameter definitions from \citet{prsa2005}.
          $\mathrm{BJD_{0}}$ refers to a previous time of superior conjunction.}
	\label{tab:phoebe_table}
	\begin{tabular}{lcr} 
		\hline
		Parameter & Prior Range & Estimate\\
		\hline
$\mathrm{P_{orb}\,[d]}$   & (4.5920,4.5930)      & $4.592834\pm3\times10^{-6}$  \\
$\mathrm{BKJD_{0}\,[JD-2454833]}$  & (2382.39,2382.418)  & $2382.4002\pm5.7\times10^{-5}$\\
$\mathrm{q}$                      & (0.05,0.3)           & $0.1723\pm0.0004$             \\
$i\, \mathrm{[\deg]}$             & (80,90)              & $83.413\pm0.006$             \\
$\mathrm{sma\,[R_{\odot}]}$       & (32,36)              & $34\pm2$                 \\
$e$                               &                      & 0 (fixed)                      \\
$\mathrm{L_1}$                    & (9.5,12.)            & $10.230\pm0.002$            \\
$\mathrm{L_2}$                    & (0.01,1.)            & $0.223\pm0.002$            \\
$\mathrm{L_3\,[\%]}$              & (0.01,0.5)           & $0.1801\pm0.0002$            \\
$\mathrm{Potential}_1\;\left(\Omega_1\right)$ & (4.5,10.0)           & $5.1006\pm0.0005$             \\
$\mathrm{Potential}_2\;\left(\Omega_2\right)$ & (3.5,10.0)           & $3.991\pm0.006$              \\
$\mathrm{T_{eff,1}\,[K]}$         &                      & $33000$ (fixed)                \\
$\mathrm{T_{eff,2}\,[K]}$         & (10000,17000)        & $12600\pm150$                \\
$\mathrm{Albedo}_1\;\left(A_1\right)$        &                      & 1 (fixed)                      \\
$\mathrm{Albedo_2}\;\left(A_2\right)$        & (0,1)                & 1 (fixed)                      \\
$f_2\,[\frac{P_{rot}}{P_{orb}}]$  & (0.1,3.)             & $1.37\pm0.101$               \\
$\mathrm{R_1\,[R_{\odot}]}$       & & $7.12\pm0.34$\\
$\mathrm{R_2\,[R_{\odot}]}$       & & $2.30\pm0.12$\\
\hline
	\end{tabular}
\newline
\footnotesize
 $\mathrm{L_1}$, $\mathrm{L_2}$, and $\mathrm{L_3}$ are scaled passband luminosities within PHOEBE.
\end{table}

\begin{figure}
\centering
\includegraphics[width=\columnwidth,keepaspectratio]{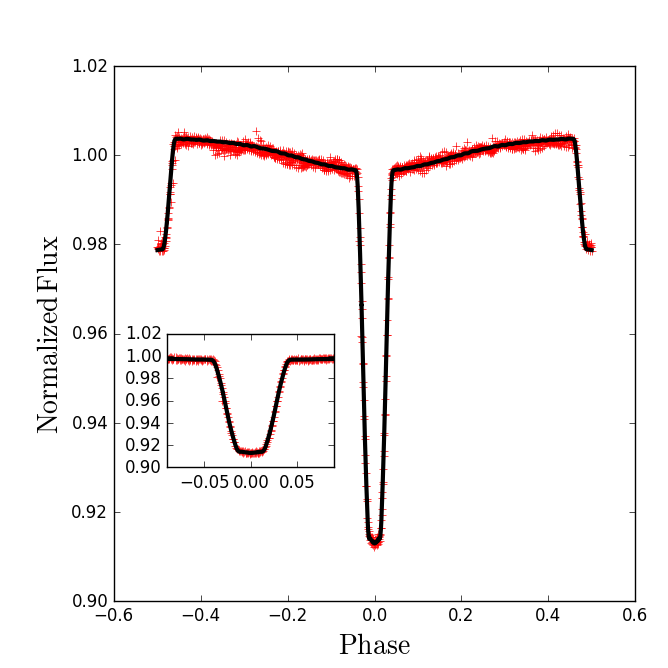}
\caption{Phase folded lightcurve of HD\,165246 in red, with the binary model in black; the inset shows a zoom in on primary eclipse.}
\label{fig:lightcurve}
\end{figure}
Utilizing the re-extracted \textit{K2\/} data, we improve upon the initial
binary solution by \cite{mayer2013}. Based on their assembled spectroscopy, we
fix the effective temperature and projected rotational velocity of the primary 
in our analysis. We model the lightcurve with the {\tt
  emcee} python module \citep{foremanMackey2013}, a Bayesian Markov Chain Monte 
  Carlo (MCMC) routine, wrapped around PHOEBE \citep{prsa2005}.

MCMC routines rely upon a Bayesian framework to evaluate the posterior
probability of a given parameter set, given a dataset, assuming a model. In our
usage the dataset is the \textit{K2\/} LC lightcurve, the model is produced by
PHOEBE, and the parameter set is comprised of the mass ratio, inclination,
semi-major axis (sma), third light, period, and reference date of the system, as well
as the potential and passband luminosity of either object, and the effective
temperature, spin/orbit synchronization, and albedo of the secondary. Our MCMC
routine evaluates 128 parameter chains for 5000 evaluations to produce 640,000
model evaluations. Initially, the values for each parameter in a chain are drawn
from uniform distributions with boundaries set as in
Table\,\ref{tab:phoebe_table}. At each iteration, each parameter chain is used
to construct a model using PHOEBE, which is then subject to a likelihood
evaluation using the following form:
\begin{equation}
    ln\mathcal{L} = -\frac{1}{2}\left( \frac{Y_i - M_{\Theta}}{\sigma_i} \right) =
-\frac{1}{2}\chi^2,
	\label{eq:likelihood}
\end{equation}
where $Y_i$ are the data, $\sigma_i$ are their uncertainties, and $M_{\Theta}$ is
the model produced with parameters $\Theta$.

PHOEBE allows for the direct interpolation of limb-darkening coefficients from pre-computed tables. Thus, we select the square-root
limb-darkening law and interpolate at each iteration. Furthermore, drawing from
the tables of \cite{claret2011}, we set the primary gravity
darkening coefficient to $\beta_1$=0.4007, assuming $T_{\rm eff,1}$=$33000\, \mathrm{K}$
and  $\mathrm{log} g_{1}$=$4.0\, \mathrm{dex}$ \citep{mayer2013}, such that the values 
adhere to the nearest values in the tables. 

The MCMC algorithm converged well within the 5000 iterations, producing Gaussian
posterior distributions for all parameters except $\mathrm{sma}$ and $A_2$. Final parameter
estimates and accompanying uncertainties were drawn from the means and standard
deviations of the posterior distributions of each parameter, as listed in
Table\,\ref{tab:phoebe_table}. The final model is shown in
Fig.\,\ref{fig:lightcurve}. In contrast to
\cite{mayer2013}, we do find a non-zero contribution of third light
($L_1$=$80.2\%, L_2$=$1.8\%, L_3$=$8\%$)
with an eccentricity of 0\textbf{, after we failed to converge to a non-zero eccentricity}. The cause of the third light cannot be unravelled due
to the large size of the \textit{Kepler\/} pixels. \cite{mayer2013} point out
that, given the near $90^{\circ}$ argument of periastron, the eccentricity that
they find might be spurious. Further long-term monitoring is required to
determine whether the eccentricity they found is real or if the orbit
is circular as we find.

Strong correlations occur between the mass ratio, $q$, and both potentials,
$\Omega_1$ and $\Omega_2$, the primary passband luminosity, $\mathrm{L_1}$, and
third light, $\mathrm{L_3}$, as well as the effective temperature and the albedo
of the secondary, $\mathrm{T_{\rm eff,2}}$ and $A_2$. The posteriors for $A_2$ 
optimized to 1, the prior boundary we set, suggesting that the value be fixed at 1.
Thus we did so, and repeated our modelling varying only $\mathrm{T_{\rm eff,2}}$,
 removing the correlation in our results. The degeneracy between the mass ratio
  and potentials cannot be avoided without further spectroscopic observations.
   The posteriors for the $\mathrm{sma}$ are nearly uniform, suggesting that there is no
information to constrain the value (within the prior range) in the data. We return to this in
Section\,\ref{sec:mesa_section} to obtain reliable estimates for the primary and
secondary masses.

\subsection{Intrinsic Variability}\label{sec:residuals_section}
\begin{figure}
\includegraphics[width=\columnwidth,keepaspectratio]{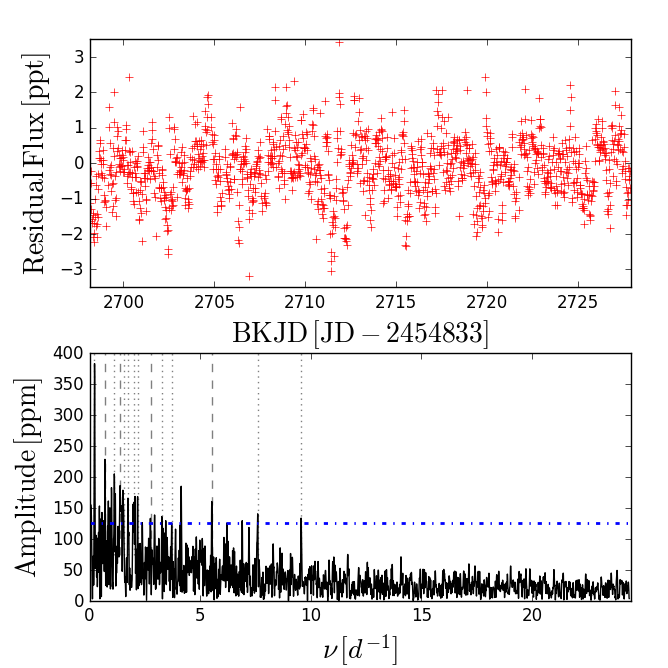}
\caption{\textbf{Top}: Residual lightcurve after removal of the binary model
  (Fig.\,\ref{fig:lightcurve}). \textbf{Bottom}: Lomb-Scargle periodogram of the
  residual lightcurve. Dotted vertical lines indicate multiples of the
  orbital frequency, while dashed lines denote multiples of the rotation
  frequency.  The horizontal line is placed at six times the average noise level after
  prewhitening all frequencies in Table\,\protect \ref{tab:freq_table}.}
    \label{fig:residuals}
\end{figure}

\begin{table}
	\centering
	\caption{Extracted frequencies, with formal statistical errors from
          non-linear least-squares fitting are given. The S/N is computed
          over $[0,24.47]\,$d$^{-1}$ after
          prewhitening. Two
          instrumental frequencies occur.}
	\label{tab:freq_table}
	\begin{tabular}{l@{\hskip 6pt}c@{\hskip 6pt}cc@{\hskip 6pt}r} 
		\hline
$\mathrm{Frequency}$ & $\mathrm{Amplitude}$ & $\mathrm{Phase}$ & $\mathrm{S/N}$ & Remarks \\
$\mathrm{[d^{-1}]}$       &  $\mathrm{[ppm]}$     & $\mathrm{modulo\,2\pi}$ & & \\
		\hline
$0.0659\pm0.0036$         &        $168\pm32$     &        $-0.39\pm0.07$         &        $8.02$   &  \\
$0.2180\pm0.0025$         &        $259\pm35$     &        $-0.03\pm0.04$         &        $12.34$  & forb \\
$0.2698\pm0.0028$         &        $251\pm39$     &        $-0.38\pm0.05$         &        $11.97$  &  \\
$0.4956\pm0.0041$         &        $131\pm28$     &        $0.15\pm0.07$          &        $6.23$   &  \\
$0.6165\pm0.0038$         &        $135\pm28$     &        $0.28\pm0.07$          &        $6.43$   &  \\
$0.6967\pm0.0027$         &        $241\pm35$     &        $-0.13\pm0.05$         &        $11.49$  & frot \\
$0.9945\pm0.0037$         &        $152\pm31$     &        $0.45\pm0.06$          &        $7.26$   &  \\
$1.1205\pm0.0031$         &        $202\pm34$     &        $-0.25\pm0.05$         &        $9.61$   & 5 forb \\
$1.3841\pm0.0034$         &        $179\pm33$     &        $0.33\pm0.06$          &        $8.52$   & 2 frot \\
$1.4247\pm0.0038$         &        $146\pm30$     &        $-0.32\pm0.07$         &        $6.96$   &  \\
$1.5090\pm0.0031$         &        $193\pm33$     &        $0.38\pm0.05$          &        $9.17$   &  \\
$1.5434\pm0.0037$         &        $147\pm29$     &        $0.13\pm0.06$          &        $6.99$   & 7 forb \\
$1.7360\pm0.0037$         &        $157\pm31$     &        $0.41\pm0.06$          &        $7.47$   & 8 forb \\
$1.9562\pm0.0040$         &        $135\pm29$     &        $-0.38\pm0.07$         &        $6.43$   &  \\
$2.0077\pm0.0036$         &        $158\pm31$     &        $0.50\pm0.06$          &        $7.51$   & 9 forb \\
$2.0544\pm0.0038$         &        $146\pm30$     &        $0.02\pm0.07$          &        $6.98$   & inst \\
$2.1706\pm0.0033$         &        $182\pm32$     &        $-0.19\pm0.06$         &        $8.68$   & 10 forb \\
$2.7544\pm0.0038$         &        $133\pm28$     &        $0.21\pm0.07$          &        $6.34$   & 4 frot \\
$2.9471\pm0.0040$         &        $129\pm28$     &        $0.13\pm0.07$          &        $6.13$   &  \\
$3.2692\pm0.0040$         &        $133\pm29$     &        $-0.12\pm0.07$         &        $6.36$   & 15 forb \\
$3.7126\pm0.0039$         &        $127\pm27$     &        $0.24\pm0.07$          &        $6.06$   & 17 forb \\
$4.1400\pm0.0033$         &        $187\pm34$     &        $-0.20\pm0.06$         &        $8.93$   & inst \\
$5.5249\pm0.0035$         &        $165\pm32$     &        $-0.43\pm0.06$         &        $7.88$   & 8 frot \\
$7.5975\pm0.0038$         &        $145\pm30$     &        $0.03\pm0.07$          &        $6.91$   & 35 forb \\
$9.5630\pm0.0039$         &        $135\pm28$     &        $-0.09\pm0.07$         &        $6.42$   & 44 forb \\
		\hline
	\end{tabular}
\end{table}
A final binary model was constructed from the values reported in
Table\,\ref{tab:phoebe_table} and removed from the lightcurve, resulting in the
top panel of Fig.\,\ref{fig:residuals}. A Lomb-Scargle periodiogram of the
residuals up to the Nyquist frequency of the LC data (24.47\,d$^{-1}$) reveals
numerous frequency peaks (bottom panel of Fig.\,\ref{fig:residuals}). The binary
model was constructed for and subtracted from the SC lightcurve as well to
confirm that no additional frequencies occur beyond 24.47\,d$^{-1}$.  

Since it has a longer time base, we proceed with the LC lightcurve to derive the
frequency content of the residuals by applying a classical prewhitening:  after
identification of the dominant frequency at each stage of the prewhitening, we
determine the frequency, amplitude, and phase using a non-linear least-squares
optimization.  Following \citet{degroote2010}, we allow the frequency to vary
within 1.5 times the Rayleigh limit, $\mathrm{R_{L}}$=$0.034\,\mathrm{d^{-1}}$ and
do not consider frequencies differing less than 0.05\,d$^{-1}$ in the
prewhitening process.  The resulting frequencies with signal-to-noise ratio
(S/N, computed over $[0,24.47]$\,d$^{-1}$) above six after prewhitening are
reported in Table\,\ref{tab:freq_table}, along with their amplitudes, phases,
and their formal statistical errors. The latter are typically a factor 10
smaller than the frequency resolution, so we rather adopt $\mathrm{R_{L}}$ as a
conservative more realistic error for the frequencies.  In this way, we identify
several frequencies as harmonics of the orbital frequency (as marked in
Table\,\ref{tab:freq_table}), likely due to the imperfect removal of the binary
signature. An asymmetry is observed between the out-of-eclipse lightcurve
approaching and receding from primary eclipse, possibly indicative of Doppler
boosting \citep{bloemen2012} and could also explain the remaining harmonics seen
in the periodogram.  Furthermore, the frequencies $f$=$\mathrm{2.0544\,d^{-1}}$
and $f$=$\mathrm{4.1409\,d^{-1}}$ occur in the spectral window and are known to 
result from the semi-periodic thruster fires of the spacecraft.

From the projected orbital velocity of the primary in the case of synchronous
rotation and the $v\sin i$ obtained from spectroscopy, \cite{mayer2013} asserted
that the synchronicity parameter of the primary is 3. Adopting this, we identify
$f$=$\mathrm{0.6967\,d^{-1}}$ as the rotation frequency.  We observe a series of even
harmonics of this frequency with decreasing amplitudes
(Table\,\ref{tab:freq_table}). This is typical for a rotational signal
intertwined with a gravity-mode oscillation spectrum in space photometry
\citep[e.g.,][]{thoul2013}.  Indeed, we interpret the remaining independent
significant frequencies in Table\,\ref{tab:freq_table} as caused by
oscillations. They can be due to heat-driven coherent modes resulting in
isolated eigenfrequencies \citep{briquet2011} or rather be connected with a
whole spectrum of frequencies as detected in three other O stars observed with
the CoRoT space mission \citep{blomme2011} and interpreted as due to IGW
\citep{aerts2015}, or a combination thereof.  The limited frequency resolution
does not allow us to distinguish among these phenomena, nor to detect frequency
or period spacing patterns. Nevertheless, we conclude that {\it the HD\,165246
  system exhibits intrinsic oscillatory variability with amplitudes of $\sim$150\,ppm level\/} and could provide a benchmark for O-star physics using complementary binary and asteroseismic modelling.

\section{Constraining Stellar Models}\label{sec:mesa_section}

Using the reported effective temperature of the primary and derived parameters
from PHOEBE, we build a set of stellar models with the MESA stellar evolution
code \citep[][and references therein,V8118]{paxton2015} with the sole aim to constrain the
masses of the binary components, while detailed modelling will be the subject of a follow-up paper based on new spectroscopy to be gathered. Our grid consists of models
with masses $\mathrm{16\,M_{\odot} \, \leq \, M \, \leq \, 25 M_{\odot}}$ in
steps of $0.5\,M_{\odot}$, adopting solar metallicity, an exponential
overshooting with parameter of $f_{ov}$=0.02, and mixing length parameter
$\alpha_{mlt}$=2.0. We apply a mass loss due to a radiation-driven
wind using an efficiency parameter $\eta_{Vink}$=0.5 in MESA \citep{vink2001}.

Models with an initial mass below $\mathrm{18.5\,M_{\odot}}$ never reach the
$3\sigma$ effective temperature range reported by \cite{mayer2013} and evolve to
lower temperatures than 32500\,K after 1.23\,Myr, leading us to dispose those. This also serves as a lower limit for the age of the system.
Given the uncertainties on the mass ratio determined through our binary
modelling, we can eliminate values of the semi-major axis below
$\mathrm{sma}$=$\mathrm{32.7\,R_{\odot}}$. Models with initial masses up to, and including,
$\mathrm{25\,M_{\odot}}$ cross the regions where the effective temperature is
consistent with spectroscopy.
 This produces a final mass range
of $\mathrm{19\,M_{\odot} \, \leq \, M_1 \, \leq \, 25 \, M_{\odot}}$ for the
primary, and range of
$\mathrm{32.7\,R_{\odot} \, \leq \, sma \, \leq \, 36 \, R_{\odot}}$ for the
semi-major axis.  Given the mass ratio obtained from binary modelling, we find a
mass range of $\mathrm{3.3\,M_{\odot} \, \leq \, M_2 \, \leq \, 4.4 \, M_{\odot}}$
for the secondary.  While the evolutionary tracks for these masses pass through
the corresponding $3-\sigma$ effective temperature found with PHOEBE, the $\log g$
is entirely unconstrained by the spectroscopy from the \citet{mayer2013}
analysis and is instead assumed from their PHOEBE model.
Due to degeneracies present in the binary modelling procedure, determination
 of the surface gravities of the components from new high S/N 
 spectroscopy is needed to break these degeneracies and further constrain theoretical models
of the system, with the potential of seismic calibration for high-mass stellar
models.  This type of data is also needed to exclude that (some of) the detected
frequencies listed in Table\,\ref{tab:freq_table} are due to the secondary or other
components.

Indeed, our mass estimate of the secondary places it in the
instability strip of slowly pulsating B stars, where gravity modes
are expected \citep{moravveji2016inst}. Although these have a narrower frequency
range and a different morphology in Fourier space than what we observe for
HD\,165246 \citep[see Fig.\,10 in][]{papics2017}, we cannot exclude that some of
the detected frequencies are due to the secondary's 1.8\% or the composite 18\%
third light contribution to the flux, as oscillation modes of amplitude above 10\,ppt do
occur in B-type pulsators. However, given reported amplitudes for IGWs in recent literature (3000 ppm,\citep{aerts2015,aerts2017}), a contaminating source would need to contribute $\sim$50\% light to the system to produce the observed amplitude, and would thus be seen in the spectra. Therefore, the IGW signal can only be produced by the hot O-type primary.

\section{Conclusions}

This work has identified HD\,165246 as the second massive O+B binary with
signatures of rotational and seismic frequencies in high-precision space
photometry.  Using 29.77 days of \textit{K2\/} data and relying on spectroscopy
in the literature, we achieved a binary model of the system. We
further detected variability in the residuals after subtraction of the binary
model and identified the rotational frequency of the O-type primary in the {\it
  K2\/} data.  The orbital frequency and several of its harmonics are still
present in the residuals, suggesting missing physics in the binary modelling
and/or additional binary variability not yet taken into account in our model.
Aside from binary and rotational variability, we detected additional frequencies
pointing towards seismic variability. The current {\it K2\/} time series is too short
to allow for an unambiguous interpretation in terms of coherent oscillation
modes or IGW.

Given the high potential of HD\,165246 to perform binary and seismic modelling,
but the low amplitude of the photometric variability, we will embark on
long-term high-precision time-series spectroscopy with various instruments and
covering several orbital periods, with the aim to clarify the nature of the
secondary and tertiary components, as well as to detect and interpret the
variability from dedicated spectral line diagnostics. Only then can we perform a full
seismic modelling of this multiple system.

\section*{Acknowledgements}
The research leading to these results has received funding from the European
Research Council (ERC) under the European Union's Horizon 2020 research and
innovation programme (grant agreement N$^\circ$670519: MAMSIE).



\bibliographystyle{mnras}
\bibliography{johnston_submitted_v3} 

\bsp	
\label{lastpage}
\end{document}